\documentclass[12pt]{article}
\usepackage{amsmath,amssymb}
\usepackage{graphicx}
\newcommand{\eqdef}{\stackrel{\text{def}}{=}}

\newcommand{\bm}{\boldsymbol}
\newcommand{\ignore}[1]{}

\newcommand{\Bxi}{\bar{\xi}}
\newcommand{\Beta}{\bar{\eta}}

\allowdisplaybreaks[4]

\begin{document}

\baselineskip=20pt

\newfont{\elevenmib}{cmmib10 scaled\magstep1}
\newcommand{\preprint}{
\vspace*{-20mm}
  \begin{flushleft}
    \elevenmib Yukawa\, Institute\, Kyoto\\
  \end{flushleft}\vspace{-1.3cm}
  \begin{flushright}\normalsize \sf
    YITP-11-24\\
  \end{flushright}}
\newcommand{\Title}[1]{{\baselineskip=26pt
  \begin{center} \Large \bf #1 \\ \ \\ \end{center}}}
\newcommand{\Author}{\begin{center}
  \large \bf Choon-Lin Ho\,${}^\ast$
  and Ryu Sasaki\,${}^\dagger$ \end{center}}
\newcommand{\Address}{\begin{center}
    $^\ast$Department of Physics, Tamkang University,
    Tamsui 251, Taiwan, R.O.C.\\
    $^\dagger$Yukawa Institute for Theoretical Physics,
    Kyoto University, Kyoto 606-8502, Japan
  \end{center}}
\newcommand{\Accepted}[1]{\begin{center}
  {\large \sf #1}\\ \vspace{1mm}{\small \sf Accepted for Publication}
  \end{center}}

\preprint \thispagestyle{empty}
\bigskip\bigskip\bigskip

\Title{Zeros of the exceptional Laguerre and Jacobi polynomials}
\Author

\Address

\date{Feb 22, 2011}

\allowdisplaybreaks[4]

\begin{abstract}

An interesting discovery in the last two years in the field of
mathematical physics has been the exceptional $X_\ell$ Laguerre
and Jacobi polynomials.  Unlike the well-known classical
orthogonal polynomials which start with constant terms, these new
polynomials have  the lowest degree $\ell=1,2,\ldots$, and yet they
form complete sets with respect to some positive-definite measure.
In this paper, we study one important aspect of these new
polynomials, namely, the behaviors of their zeros as some
parameters of the Hamiltonians change.

\end{abstract}





\section{Introduction}

The discovery of new types of orthogonal polynomials, called the
exceptional $X_\ell$ polynomials, has been the most interesting
development in the area of exactly solvable models in quantum
mechanics in the last two years \cite{GKM1,Que,OS1,HOS,DR}. Unlike
the classical orthogonal polynomials, these new polynomials have
the remarkable properties that they still form complete sets with
respect to some positive-definite measure, although they start
with degree $\ell$ polynomials instead of a constant.  Four sets
of infinite families of such polynomials, namely, the Laguerre
type L1,~L2, and the Jacobi type J1, J2 $X_\ell$ polynomials, with
$\ell=1,2,\ldots$, were constructed in \cite{OS1}. These systems
were derived by deforming the radial oscillator potential and the
Darboux-P\"oschl-Teller (DPT) potential in terms of an eigenpolynomial
of degree $\ell$ ($\ell=1,2,\ldots$).  The lowest ($\ell=1$)
examples, the $X_1$-Laguerre and $X_1$-Jacobi polynomials, are
equivalent to those introduced in the pioneering work of
Gomez-Ullate et al. \cite{GKM1} within the Sturm-Liouville theory.
The results in \cite{GKM1} were reformulated in the framework of
quantum mechanics and shape-invariant potentials by Quesne et al.
\cite{Que}. By construction these new orthogonal polynomials
satisfy a second order differential equation (the Schr\"odinger
equation) without contradicting Bochner's theorem \cite{Bochner},
since they start at degree $\ell>0$ instead of the degree zero
constant term.  Generalization of exceptional orthogonal
polynomials to discrete quantum mechanical systems was done in
\cite{OS2}.

Later, equivalent but much simpler looking forms of the Laguerre-
and Jacobi-type $X_{\ell}$ polynomials than those originally
presented in \cite{OS1} were given in \cite{HOS}.  These nice
forms were derived based on an analysis of the second order
differential equations for the $X_{\ell}$ polynomials within the
framework of the Fuchsian differential equations in the entire
complex $x$-plane.  They allow us to study in-depth some important
properties of the $X_\ell$ polynomials, such as the actions of the
forward and backward shift operators on the $X_{\ell}$
polynomials, Gram-Schmidt orthonormalization for the algebraic
construction of the $X_{\ell}$ polynomials, Rodrigues formulas,
and the generating functions of these new polynomials.

Recently, these exceptional orthogonal polynomials were generated by
means of the Darboux-Crum transformation \cite{GKM2,STZ}. Physical
models which may involve these new polynomials were considered in
\cite{Ho}.

One important aspect related to these new polynomials, which was
only briefly mentioned in \cite{HOS} but has not been investigated
in-depth so far, is the structure of their zeros. It is the
purpose of this paper to look into this. Particularly, we
investigate the behaviors of the zeros as the parameters of the
polynomials change.

The plan of this paper is as follows.  In Sect.~2 we briefly
review the forms of the exceptional polynomials. Sect.~3 and 4
study the behaviors of the extra and the ordinary zeros, respectively,
of the exceptional polynomials as one of $\ell$ and $n$ increases
while the other parameters being kept fixed.  Sect.~5 presents
analytical proofs that explain the movements of the extra zeros of
the exceptional polynomials as $n$ changes at fixed $\ell$.  In
Sect.~6 we consider behaviors of the zeros at large $g$ and/or
$h$.  Sect.~7 summarizes the paper.

\section{Exceptional orthogonal polynomials}

Four sets of infinitely many exceptional orthogonal polynomials
were derived in \cite{OS1}, among them two are deformations of the
Laguerre polynomials, and the others are deformations of the
Jacobi polynomials.  A unified nice form of these polynomials was
given in \cite{HOS}, in which these polynomials are expressed as a
bilinear form of the original polynomials, the Laguerre or Jacobi
polynomials and the deforming polynomials, depending on the set of
parameters $\bm{\lambda}$ and their shifts $\bm{\delta}$ and a
non-negative integer $\ell$, which is the degree of the deforming
polynomials. The two sets of exceptional Laguerre polynomials
($\ell=1,2,\ldots$, $n=0,1,2,\ldots$) are:
\begin{equation}
 P_{\ell,n}(\eta;\bm{\lambda})\eqdef
 \left\{
 \begin{array}{ll}
 \xi_{\ell}(\eta;\bm{\lambda}+\bm{\delta})P_n(\eta;g+\ell-1)
 -\xi_{\ell}(\eta;\bm{\lambda})\partial_{\eta}
 P_n(\eta;g+\ell-1)&:\text{L1}\\[2pt]
 (g+\frac12)
 \xi_{\ell}(\eta;\bm{\lambda}+\bm{\delta})P_n(\eta;g+\ell+1)\\
 \phantom{(n+g+\frac12)^{-1}\bigl(}\ \quad
 +\eta\xi_{\ell}(\eta;\bm{\lambda})\partial_{\eta}P_n(\eta;g+\ell+1)
&:\text{L2}
 \end{array}\right.,
 \label{XLform}
\end{equation}
in which $\bm{\lambda}\eqdef g>0$ and $\bm{\delta}\eqdef 1$ and
\begin{equation}
 P_n(\eta;g)\eqdef L_n^{(g-\frac{1}{2})}(\eta),\quad
 \xi_{\ell}(\eta;g)\eqdef
 \left\{
 \begin{array}{ll}
 L_{\ell}^{(g+\ell-\frac32)}(-\eta)&:\text{L1}\\
 L_{\ell}^{(-g-\ell-\frac12)}(\eta)&:\text{L2}
 \end{array}\right..
 \label{defxiL}
\end{equation}
The two sets of exceptional Jacobi polynomials ($\ell=1,2,\ldots$,
$n=0,1,2,\ldots$) are :
\begin{equation}
 P_{\ell,n}(\eta;\bm{\lambda})\eqdef
 \left\{
 \begin{array}{ll}
 \!\!\!
 (h+\frac12)\xi_{\ell}(\eta;\bm{\lambda}+\bm{\delta})
 P_n(\eta;g+\ell-1,h+\ell+1)&\\
 \!\!\!
 \phantom{(n+h+\frac12)^{-1}\bigl(}
 +(1+\eta)\xi_{\ell}(\eta;\bm{\lambda})
 \partial_{\eta}P_{n}(\eta;g+\ell-1,h+\ell+1)
 &\!\!\!:\text{J1}\\[2pt]
 \!\!\!
 (g+\frac12)\xi_{\ell}(\eta;\bm{\lambda}+\bm{\delta})
 P_n(\eta;g+\ell+1,h+\ell-1)&\\
 \!\!\!
 \phantom{(n+g+\frac12)^{-1}\bigl(}
 -(1-\eta)\xi_{\ell}(\eta;\bm{\lambda})
 \partial_{\eta}P_{n}(\eta;g+\ell+1,h+\ell-1)
 &\!\!\!:\text{J2}
 \end{array}\right.,
 \label{XJform}
\end{equation}
in which $\bm{\lambda}\eqdef(g,h)$, $g>0$, $h>0$,
$\bm{\delta}\eqdef(1,1)$ and
\begin{equation}
 P_n(\eta;g,h)\eqdef P_n^{(g-\frac12,h-\frac12)}(\eta),\ \
 \xi_{\ell}(\eta;g,h)\eqdef
 \left\{
 \begin{array}{ll}
 \!\!\!
 P_{\ell}^{(g+\ell-\frac32,-h-\ell-\frac12)}(\eta),\ \ g>h>0&\!\!:\text{J1}\\
 \!\!\!
 P_{\ell}^{(-g-\ell-\frac12,h+\ell-\frac32)}(\eta),\ \ h>g>0&\!\!:\text{J2}
 \end{array}\right..
 \label{defxiJ}
\end{equation}
The new exceptional orthogonal polynomials can be viewed as
deformations of the classical orthogonal polynomials by the
parameter $\ell$, and the two polynomials
$\xi_{\ell}(\eta;\bm{\lambda})$ and
$\xi_{\ell}(\eta;\bm{\lambda}+\bm{\delta})$ played the role of the
deforming polunomials.

The zeros of orthogonal polynomials have always attracted the
interest of researchers.  In this paper we shall study the
properties of the zeros of these new exceptional polynomials as
some of their basic parameters change.

In the case of $X_{\ell}$ polynomial
$P_{\ell,n}(\eta;\bm{\lambda})$, it has $n$ zeros in the
(ordinary) domain where the weight function is defined, that is
$(0,\infty)$ for the L1 and L2 polynomials and $(-1,1)$ for the J1
and J2 polynomials. The behavior of these zeros, which we shall
call the ordinary zeros, are the same as those of other ordinary
orthogonal polynomials. We shall say more about these zeros in
sect.~\ref{ordinary-zeros}. Besides these $n$ zeros, there are
extra $\ell$ zeros outside the ordinary domain. For convenience,
we shall adopt the following notation for the zeros of the various
polynomials involved:
\begin{align}
\Bxi^{(\ell)}_k:&~  {\rm zeros\  of}~~
\xi_{\ell}(\eta;\bm{\lambda}+\bm{\delta}),  & k=1,2,\ldots,\ell;\\
\xi^{(\ell)}_k: &~ {\rm zeros\  of}~~
\xi_{\ell}(\eta;\bm{\lambda}), & k=1,2,\ldots,\ell;\\
\Beta_k^{(\ell,n)}:&~ {\rm extra\ zeros\  of}~~ P_{\ell,n},
&k=1,2,\ldots,\ell;
\\
\eta_j^{(\ell,n)}:&~ {\rm ordinary\ zeros\  of}~~ P_{\ell,n}, &j
=1,2,\ldots,n.
\end{align}
We emphasize that $\eta_j^{(\ell,n)}\in (0,\infty)$ for the L1 and
L2 Laguerre polynomials, and $\eta_j^{(\ell,n)}\in (-1,1)$ for the
J1, and J2 Jacobi polynomials.

Figs.~1--10 depict the distribution of the zeros for some
representative parameters of the systems, namely, $n, \ell, g$ and
$h$.  From these figures one can deduce certain patterns of the
distribution of the zeros as those parameters vary.  We will
discuss these behaviors below.

\section{Extra $\ell$ zeros of $X_{\ell}$ polynomials}

Here we discuss the location of the extra $\ell$ zeros of the
exceptional orthogonal polynomials, which lie in various different
positions for the different types of polynomials. From our
numerical analysis, we can summarize the trend as follows.

The $\ell$ extra zeros of L1 polynomials are on the negative real
line $(-\infty,0)$. The L2 $X_{\ell \text{:odd}}$ polynomials have
one real negative zero which lies to the left of the remaining
$\frac12(\ell-1)$ pairs of complex conjugate roots. The L2
$X_{\ell \text{:even}}$ polynomials have $\frac12\ell$ pairs of
complex conjugate roots.

The situations for the $X_\ell$ Jacobi polynomials are a bit more
complicated. The J1 $X_{\ell \text{:odd}}$ polynomials have one real
negative root which lies to the left of the remaining
$\frac12(\ell-1)$  pairs of complex conjugate roots with negative
real parts. The J1 $X_{\ell \text{:even}}$ polynomials have
$\frac12\ell$ pairs of complex conjugate roots with negative  real
parts. The J2 $X_{\ell \text{:odd}}$ polynomials have one real
positive root which lies to the right of the remaining
$\frac12(\ell-1)$  pairs of complex conjugate roots with positive
real parts. The J2 $X_{\ell \text{:even}}$ polynomials have
$\frac12\ell$ pairs of complex conjugate roots with positive  real
parts.

One notes that the J1 and J2 polynomials are the mirror images of
each other, in the sense $\eta\leftrightarrow-\eta$ and
$g\leftrightarrow h$, as exemplified by the relation
$\xi_{\ell}^{\text{J2}}(\eta;g,h)=(-1)^\ell\xi_{\ell}^{\text{J1}}(-\eta;h,g)$
\cite{OS1,HOS}.  So the behaviors of the zeros of J1 Jacobi
polynomials can be obtained from those of the J2 type accordingly.
As such, for clarity of presentation, we shall only discuss the
behaviors of the zeros of the J2 Jacobi polynomials in this paper.

\subsection{Behaviors as $n$ increases at fixed $\ell$}

In all cases, we have
\begin{equation}
P_{\ell,0}(\eta;\bm{\lambda})\propto
\xi_{\ell}(\eta;\bm{\lambda}+\bm{\delta}).
\end{equation}
This implies that the zeros of $P_{\ell,0}$ coincide with those of
$\xi_{\ell}(\eta;\bm{\lambda}+\bm{\delta})$, namely,
$\Bxi^{(\ell)}_k,~k=1,2,\ldots, \ell$.

\begin{figure}[htbp]
\caption{\label{fig1} L1: Distributions of the zeros
$\Beta_k^{(\ell,n)},~\eta_j^{(\ell,n)}$ ($\blacklozenge$),
$\Bxi^{(\ell)}_k$ ($\bigcirc$) and $\xi^{(\ell)}_k $
($\blacksquare$) for the L1 Laguerre polynomials, with $g=0.5$ and
$\ell=2$.  The three diagrams correspond to $n=1 (a), 2 (b)$ and
$3 (c)$, respectively. The ordinary zeros $\eta_j^{(\ell,n)}$ lie
in $(0,\infty)$.}
\begin{center}
(a) ~~~ \includegraphics{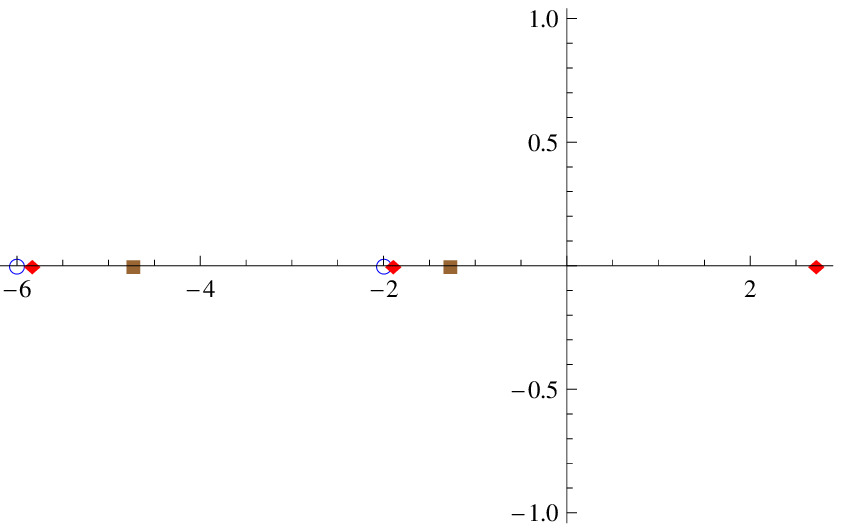}\\
~\\
(b) ~~~ \includegraphics{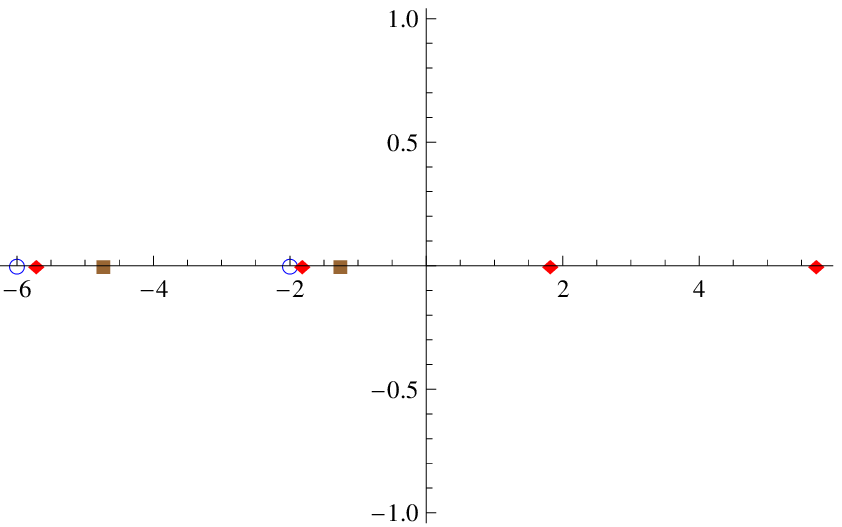}\\
~\\
(c) ~~~ \includegraphics{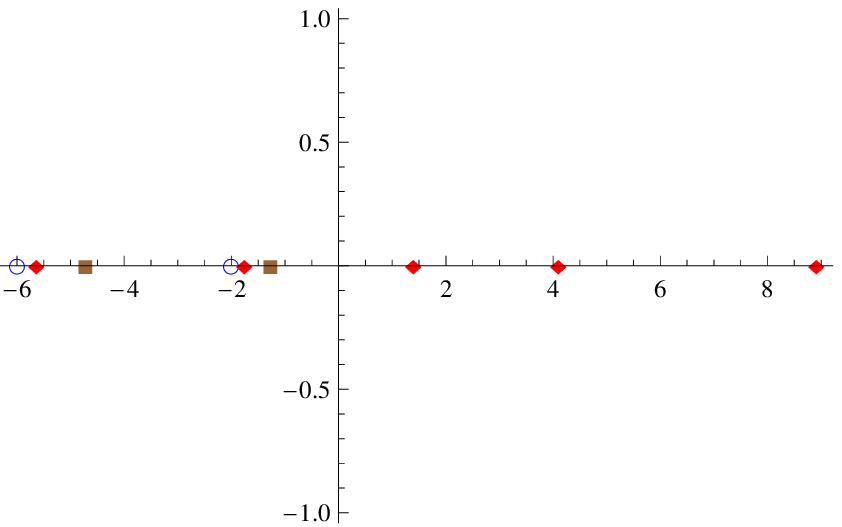}
\end{center}
\end{figure}

\begin{figure}[htbp]
\caption{\label{fig2} L2: Distributions of the zeros
$\Beta_k^{(\ell,n)},~\eta_j^{(\ell,n)}$ ($\blacklozenge$),
$\Bxi^{(\ell)}_k$ ($\bigcirc$) and $\xi^{(\ell)}_k $
($\blacksquare$) for the L2 Laguerre polynomials, with $g=0.5$ and
$\ell=3$.  The three diagrams correspond to $n=1 (a), 2 (b)$ and
$5 (c)$, respectively. The ordinary zeros $\eta_j^{(\ell,n)}$ lie
in $(0,\infty)$.}
\begin{center}
(a) ~~~ \includegraphics{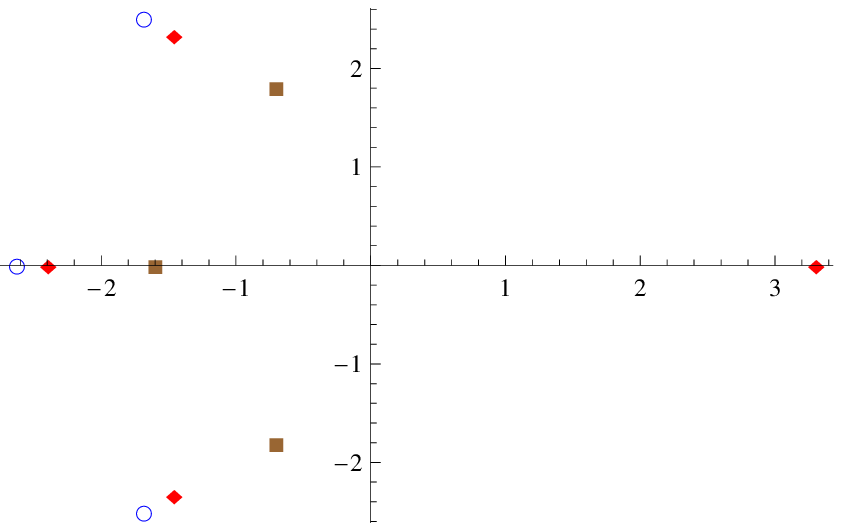}\\
~\\
(b) ~~~ \includegraphics{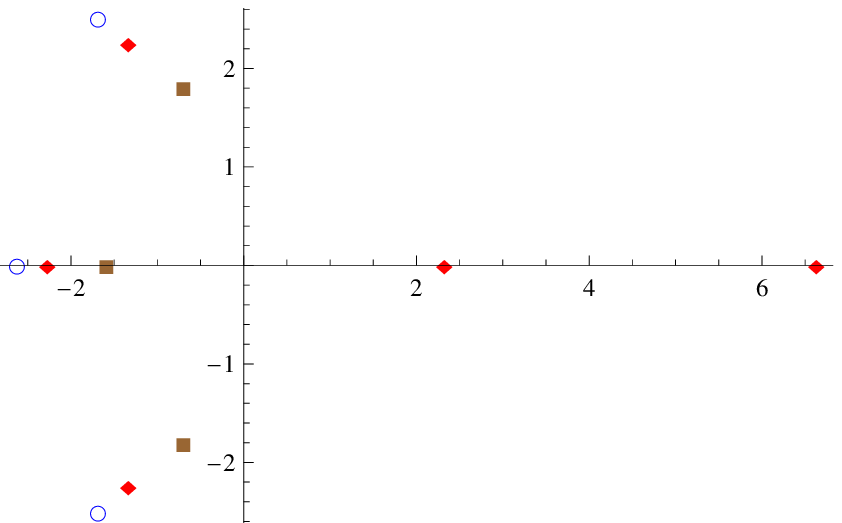}\\
~\\
(c) ~~~ \includegraphics{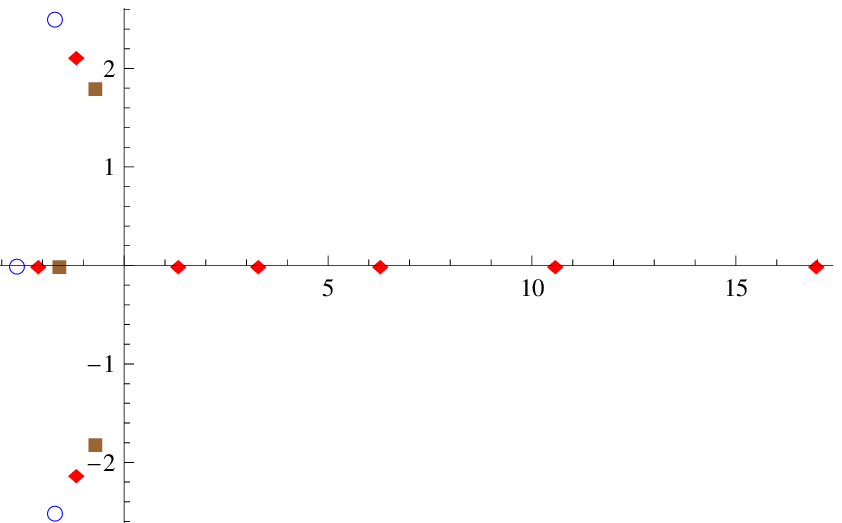}
\end{center}
\end{figure}

\begin{figure}[htbp]
\caption{\label{fig3} J2: Distributions of the zeros
$\Beta_k^{(\ell,n)},~\eta_j^{(\ell,n)}$ ($\blacklozenge$),
$\Bxi^{(\ell)}_k$ ($\bigcirc$) and $\xi^{(\ell)}_k $
($\blacksquare$) for the J2 Jacobi polynomials, with $g=3, h=4$
and $\ell=3$.  The three diagrams correspond to $n=1 (a), 2 (b)$
and $5 (c)$, respectively. The ordinary zeros $\eta_j^{(\ell,n)}$
lie in $(-1,1)$.}
\begin{center}
(a) ~~~ \includegraphics{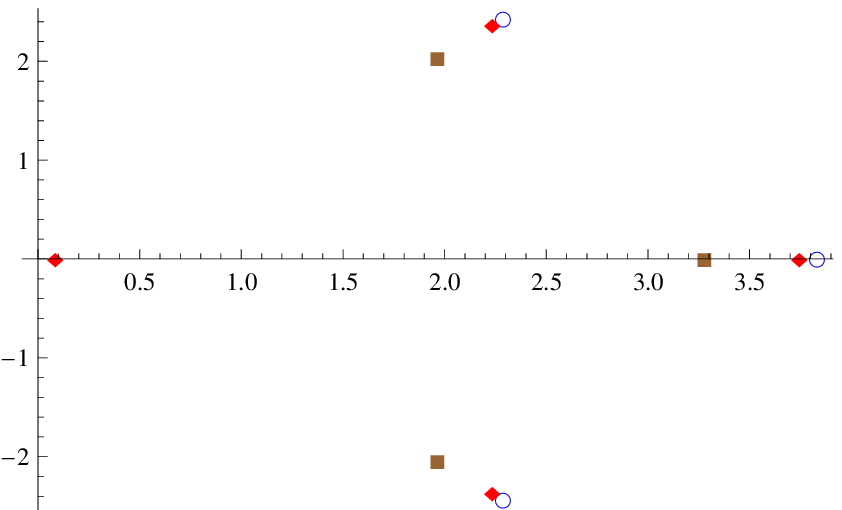}\\
~\\
(b) ~~~ \includegraphics{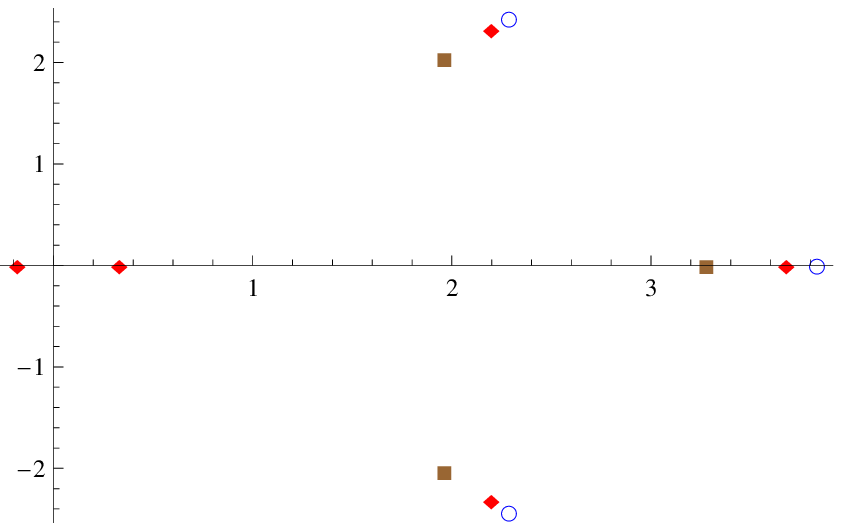}\\
~\\
(c) ~~~ \includegraphics{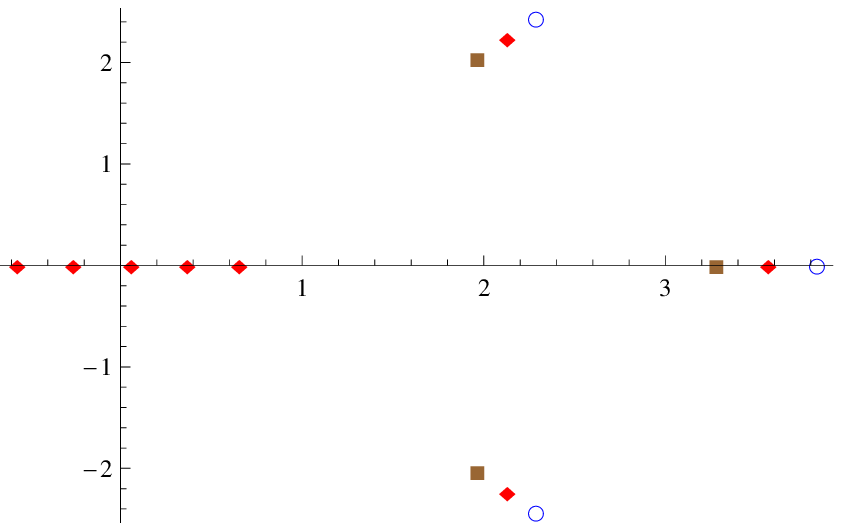}
\end{center}
\end{figure}

At fixed $\ell$, all the $\Beta^{(\ell,n)}_k$ move from
$\Bxi^{(\ell)}_k$ at $n=0$, to $\xi^{(\ell)}_k$ as $n\to \infty$.
This can be seen from Figs. \ref{fig1}--\ref{fig3} and in Tables
\ref{table1}--\ref{table7}. We shall prove this result generally
in Sect.~\ref{sect:eta-flow}.


\begin{table}[!]
\caption{\label{table1} List of the zeros $\Bxi^{(\ell)}_k$,
$\xi^{(\ell)}_k $ and $\Beta_k^{(\ell,n)}$ for the L1 Laguerre
polynomials with $g=2$, $\ell=5$, and $n=0,10,20,\ldots,60$
($k=1,2\ldots,\ell$). It can be seen that when $n=0$,
$\Beta_k^{(\ell,n=0)}=\Bxi^{(\ell)}_k$. As $n$ increases,
$\Beta_k^{(\ell,n)}$ approach to $\xi^{(\ell)}_k $.}
\begin{center}
\begin{tabular}{rrrrrr}
\hline\hline
$\Bxi^{(\ell)}_k:$ ~~~~~~~ &
-22.4802 & -15.2391 & -10.1403 & -6.2977 & -3.3427\\
\hline $n=0$  &
-22.4802 & -15.2391 & -10.1403 & -6.2977 & -3.3427\\
 10   &
-22.0686 & -14.8767 &  -9.8314 & -6.0505 & -3.1698\\
 20   &
-21.8830 & -14.7189 &  -9.7004 & -5.9469 & -3.0962\\
$\Beta_k^{(\ell,n)}$: ~~  30   &
-21.7717 & -14.6253 &  -9.6233 & -5.8862 & -3.0529\\
 40   &
-21.6954 & -14.5617 &  -9.5711 & -5.8452 & -3.0237\\
 50   &
-21.6390 & -14.5148 &  -9.5327 & -5.8152 & -3.0022\\
 60   &
-21.5951 & -14.4784 &  -9.5030 & -5.7919 & -2.9856\\
\hline $\xi^{(\ell)}_k:$ ~~~~~~~ &
-21.0456 & -14.0274 & -9.1375 & -5.5071 & -2.7824\\
\hline\hline
\end{tabular}
\end{center}
\end{table}

\begin{table}[!]
\begin{center}
\caption{\label{table2} Same as Table 1 for L1 Laguerre
polynomials with $g=8$ and $\ell=5$.}
\smallskip
\begin{tabular}{rrrrrr}
\hline\hline
$\Bxi^{(\ell)}_k:$ ~~~~~~~ &
-30.7592 & -22.3415 & -16.1499 & -11.2032 & -7.0462\\
\hline
$n=0$ &
-30.7592 & -22.3415 & -16.1499 & -11.2032 & -7.0462\\
 10  &
-30.4724 & -22.0859 & -15.9269 & -11.0165 & -6.9029\\
 20  &
-30.3144 & -21.9474 & -15.8074 & -10.9169 & -6.8255\\
$\Beta_k^{(\ell,n)}$: ~~  30  &
-30.2107 & -21.8574 & -15.7301 & -10.8525 & -6.7752\\
 40  &
-30.1361 & -21.7928 & -15.6748 & -10.8065 & -6.7393\\
 50  &
-30.0791 & -21.7436 & -15.6328 & -10.7715 & -6.7119\\
 60  &
-30.0336 & -21.7046 & -15.5994 & -10.7438 & -6.6902\\
\hline
$\xi^{(\ell)}_k:$ ~~~~~~~ &
-29.4106 & -21.1735 & -15.1488 & -10.3703 & -6.3968\\
\hline\hline
\end{tabular}
\end{center}
\end{table}

\begin{table}[!]
\begin{center}
\caption{\label{table3} Same as Table 1 but for L2 Laguerre
polynomials with $g=3$ and $\ell=4$.}
\smallskip
\begin{tabular}{rrr}
\hline\hline
$\Bxi^{(\ell)}_k:$ ~~~~~~~
     & -5.29007 $\pm$ 1.65310 $i$ & -3.70993 $\pm$ 5.05130 $i$\\
\hline
$n=0$ & -5.29007 $\pm$ 1.65310 $i$ & -3.70993 $\pm$ 5.05130 $i$\\
 10  & -4.84198 $\pm$ 1.57129 $i$ & -3.25524 $\pm$ 4.78004 $i$\\
 20  & -4.71299 $\pm$ 1.54888 $i$ & -3.12839 $\pm$ 4.70776 $i$\\
 $\Beta_k^{(\ell,n)}$: ~~
 30  & -4.64523 $\pm$ 1.53732 $i$ & -3.06246 $\pm$ 4.67065 $i$\\
 40  & -4.60183 $\pm$ 1.52998 $i$ & -3.02046 $\pm$ 4.64713 $i$\\
 50  & -4.57100 $\pm$ 1.52479 $i$ & -2.99074 $\pm$ 4.63053 $i$\\
 60  & -4.54766 $\pm$ 1.52087 $i$ & -2.96828 $\pm$ 4.61801 $i$\\
\hline
$\xi^{(\ell)}_k:$ ~~~~~~~
     & -4.28361 $\pm$ 1.47684 $i$ & -2.71639 $\pm$ 4.47739 $i$\\
\hline\hline
\end{tabular}
\end{center}
\end{table}

\begin{table}[!]
\begin{center}
\caption{\label{table4} Same as Table 3 for L2 Laguerre
polynomials with $g=10$ and $\ell=5$.}
\smallskip
\begin{tabular}{rrrr}
\hline\hline
$\Bxi^{(\ell)}_k:$ ~~~~~~~
      & -12.8111 & -12.2115 $\pm$ 4.7185 $i$ & -10.1329 $\pm$ 9.7965 $i$\\
\hline
$n=0$ &  -12.8111 & -12.2115 $\pm$ 4.7185 $i$ & -10.1329 $\pm$ 9.7965 $i$\\
   10 &  -12.5476 & -11.9465 $\pm$ 4.6639 $i$ &  -9.8622 $\pm$ 9.6780 $i$\\
   20 &  -12.4210 & -11.8198 $\pm$ 4.6384 $i$ &  -9.7348 $\pm$ 9.6233 $i$\\
$\Beta_k^{(\ell,n)}$: ~~
   30 &  -12.3430 & -11.7418 $\pm$ 4.6229 $i$ &  -9.6570 $\pm$ 9.5901 $i$\\
   40 &  -12.2888 & -11.6877 $\pm$ 4.6122 $i$ &  -9.6032 $\pm$ 9.5673 $i$\\
   50 &  -12.2483 & -11.6473 $\pm$ 4.6043 $i$ &  -9.5632 $\pm$ 9.5504 $i$\\
   60 &  -12.2165 & -11.6157 $\pm$ 4.5981 $i$ &  -9.5319 $\pm$ 9.5372 $i$\\
\hline
$\xi^{(\ell)}_k:$ ~~~~~~~
      & -11.8092 & -11.2107 $\pm$ 4.5195 $i$ &  -9.1347 $\pm$ 9.3702 $i$\\
\hline\hline
\end{tabular}
\end{center}
\end{table}

\begin{table}[!]
\begin{center}
\caption{\label{table5} Same as Table 1 but for J2 Jacobi
polynomials with $g=3,~h=4$ and $\ell=4$.}
\smallskip
\begin{tabular}{rcc}
\hline\hline
$\Bxi^{(\ell)}_k:$ ~~~~~~~
      & 1.56846 $\pm$ 2.10278 $i$ & 3.00297 $\pm$ 0.91199 $i$\\
\hline
$n=0$ & 1.56846 $\pm$ 2.10278 $i$ & 3.00297 $\pm$ 0.91199 $i$ \\
  10  & 1.45201 $\pm$ 1.89890 $i$ & 2.76834 $\pm$ 0.82626 $i$ \\
  20  & 1.42407 $\pm$ 1.85433 $i$ & 2.71360 $\pm$ 0.80733 $i$ \\
$\Beta_k^{(\ell,n)}$: ~~
  30  & 1.41139 $\pm$ 1.83435 $i$ & 2.68882 $\pm$ 0.79884 $i$ \\
  40  & 1.40414 $\pm$ 1.82297 $i$ & 2.67466 $\pm$ 0.79401 $i$ \\
  50  & 1.39944 $\pm$ 1.81561 $i$ & 2.66550 $\pm$ 0.79088 $i$ \\
  60  & 1.39615 $\pm$ 1.81046 $i$ & 2.65907 $\pm$ 0.78869 $i$ \\
\hline
$\xi^{(\ell)}_k:$ ~~~~~~~
      & 1.37745 $\pm$ 1.78118 $i$ & 2.62255 $\pm$ 0.77624 $i$\\
\hline\hline
\end{tabular}

\end{center}
\end{table}

\begin{table}[!]
\begin{center}
\caption{\label{table6} Same as Table 5 for J2 Jacobi polynomials
with $g=3,~h=4$ and $\ell=5$.}
\smallskip
\begin{tabular}{rccc}
\hline\hline
$\Bxi^{(\ell)}_k:$  & 1.19188 $\pm$ 1.85256 $i$ & 2.38851 $\pm$ 1.21416 $i$ & 2.83923 \\
\hline
$n=0$ & 1.19188 $\pm$ 1.85256 $i$ & 2.38851 $\pm$ 1.21416 $i$ & 2.83923\\
  10  & 1.11856 $\pm$ 1.68660 $i$ & 2.22979 $\pm$ 1.11021 $i$ & 2.64753\\
  20  & 1.09936 $\pm$ 1.64851 $i$ & 2.18998 $\pm$ 1.08600 $i$ & 2.59983\\
$\Beta_k^{(\ell,n)}$: ~~
  30  & 1.09041 $\pm$ 1.63110 $i$ & 2.17151 $\pm$ 1.07491 $i$ & 2.57771\\
  40  & 1.08522 $\pm$ 1.62106 $i$ & 2.16081 $\pm$ 1.06852 $i$ & 2.56490\\
  50  & 1.08184 $\pm$ 1.61453 $i$ & 2.15382 $\pm$ 1.06436 $i$ & 2.55654\\
  60  & 1.07945 $\pm$ 1.60993 $i$ & 2.14890 $\pm$ 1.06143 $i$ & 2.55066\\
\hline
$\xi^{(\ell)}_k:$  & 1.06566 $\pm$ 1.58339 $i$ & 2.12047 $\pm$ 1.04452 $i$ & 2.51663\\
\hline\hline
\end{tabular}
\end{center}
\end{table}

\begin{table}[!]
\begin{center}
\caption{\label{table7} same as Table 5 for J2 Jacobi polynomials
with $g=8,~h=9$ and $\ell=3$.}
\smallskip
\begin{tabular}{rrr}
\hline\hline
$\Bxi^{(\ell)}_k:$ ~~~~~~~
     & 3.90615 $\pm$ 4.35051 $i$ & 6.58770\\
\hline
$n=0$ & 3.90615 $\pm$ 4.35051 $i$ & 6.58770\\
 10  & 3.74981 $\pm$ 4.16635 $i$ & 6.32188\\
 20  & 3.69527 $\pm$ 4.10323 $i$ & 6.22948\\
$\Beta_k^{(\ell,n)}$: ~~
 30  & 3.66745 $\pm$ 4.07116 $i$ & 6.18240\\
 40  & 3.65057 $\pm$ 4.05174 $i$ & 6.15383\\
 50  & 3.63924 $\pm$ 4.03870 $i$ & 6.13465\\
 60  & 3.63110 $\pm$ 4.02935 $i$ & 6.12088\\
\hline
$\xi^{(\ell)}_k:$ ~~~~~~~
     & 3.58151 $\pm$ 3.97238 $i$ & 6.03699\\
\hline\hline
\end{tabular}
\end{center}
\end{table}

\subsection{Behaviors as $\ell$ increases at fixed $n$}

The discussions in the last subsection show that
$\Beta^{(\ell,n)}_k$ are sandwiched between $\Bxi^{(\ell)}_k$ and
$\xi^{(\ell)}_k$. Thus to know how $\Beta^{(\ell,n)}_k$ behave as
$\ell$ increases at fixed $n$, we only need to study how the zeros
$\Bxi^{(\ell)}_k$ and $\xi^{(\ell)}_k$ flow as $\ell$ increases.

\subsubsection{L1 Laguerre}

As $\ell$ changes to $\ell+1$, the zeros of $\xi_\ell(\eta; g+1)$ and
$\xi_\ell(\eta; g)$ decrease (move to the left), and a new set of zeros
appear from the right.

$\Bxi^{(\ell +1)}_k < \Bxi^{(\ell) }_k ,$

$\xi^{(\ell +1)}_k < \xi^{(\ell) }_k, $

$\Bxi^{(\ell) }_k <  \xi^{(\ell)}_k < \Bxi^{(\ell) }_{k+1} <
\xi^{(\ell) }_{k+1}, $

for $~~k=1,2,\ldots,\ell-1,\ell$.

\begin{figure}[htbp]
\caption{\label{fig4} L1: Distributions of the zeros
$\Beta_k^{(\ell,n)},~\eta_j^{(\ell,n)}$ ($\blacklozenge$),
$\Bxi^{(\ell)}_k$ ($\bigcirc$) and $\xi^{(\ell)}_k $
($\blacksquare$) for the L1 Laguerre polynomials, with $g=0.5$ and
$n=2$.  The three diagrams correspond to $\ell=1 (a), 2 (b)$ and
$3 (c)$, respectively.}
\begin{center}
(a) ~~~ \includegraphics{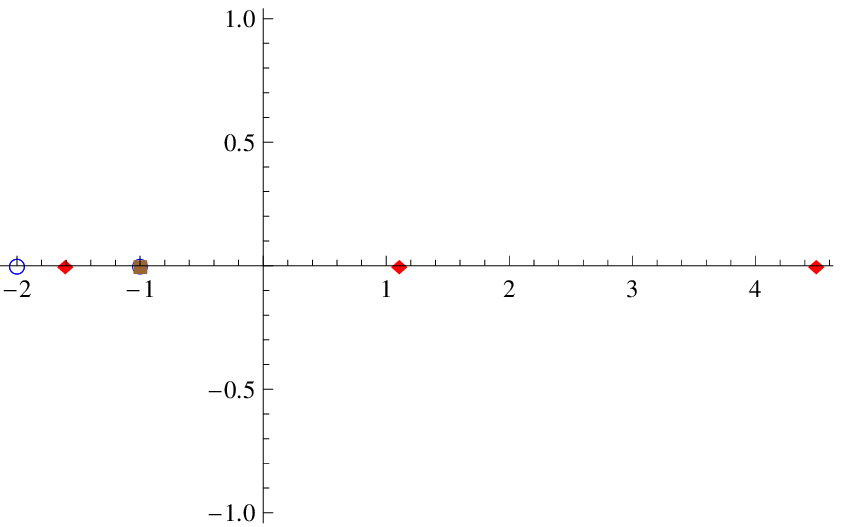}\\
~\\
(b) ~~~ \includegraphics{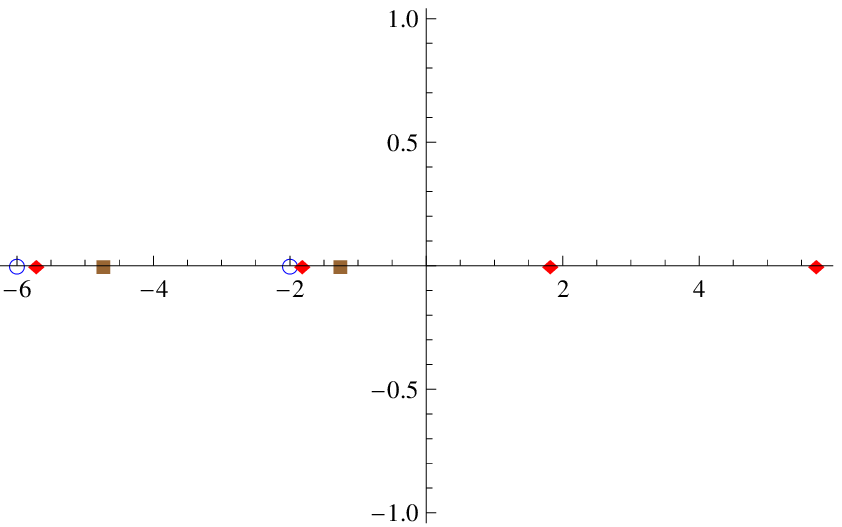}\\
~\\
(c) ~~~ \includegraphics{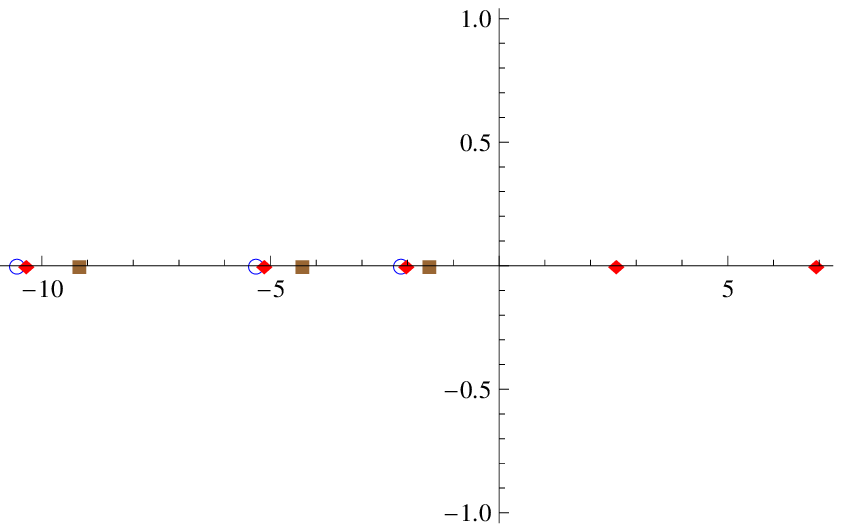}
\end{center}
\end{figure}

\begin{figure}[htbp]
\caption{\label{fig5} L1: Distributions of the zeros
$\Beta_k^{(\ell,n)},~\eta_j^{(\ell,n)}$ ($\blacklozenge$),
$\Bxi^{(\ell)}_k$ ($\bigcirc$) and $\xi^{(\ell)}_k $
($\blacksquare$) for the L1 Laguerre polynomials, with $g=1.5$ and
$n=2$.  The three diagrams correspond to $\ell=1 (a), 2 (b)$ and
$3 (c)$, respectively.}
\begin{center}
(a) ~~~ \includegraphics{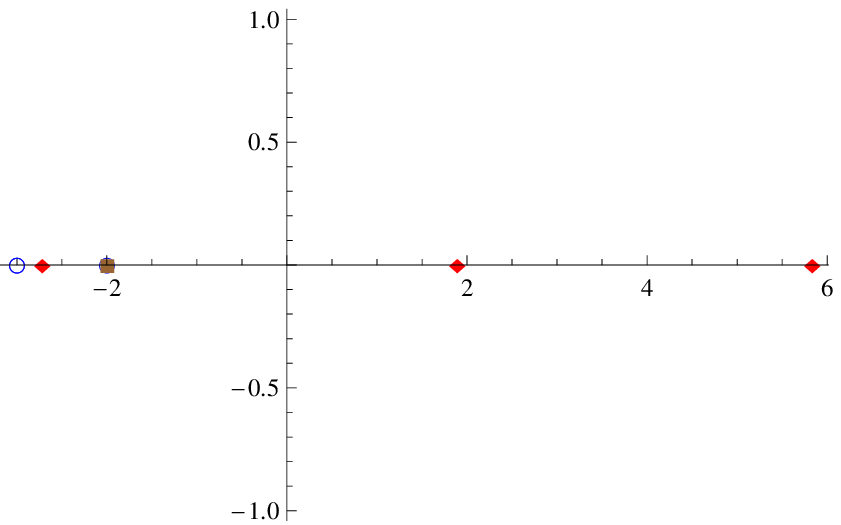}\\~\\
(b) ~~~ \includegraphics{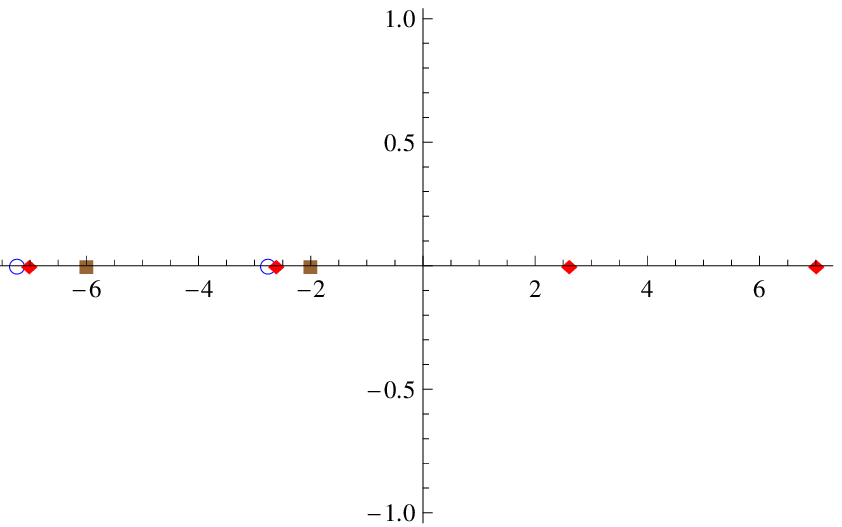}\\~\\
(c) ~~~ \includegraphics{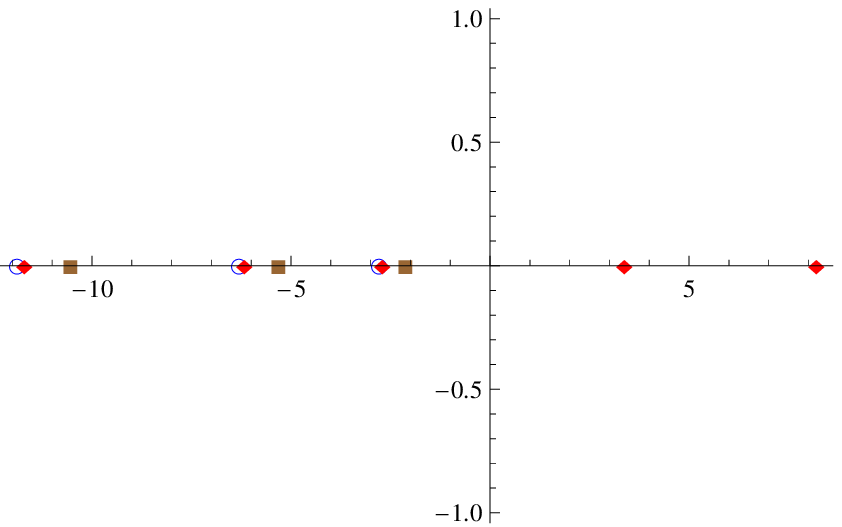}
\end{center}
\end{figure}

We show these patterns for some representative parameters in
Figs.~\ref{fig4} and \ref{fig5}.

\subsubsection{L2 Laguerre}

For $\ell=1$, there is one real root each for $\xi_\ell(\eta;
g+1)$ and $\xi_\ell(\eta; g)$,
with $\Bxi^{(\ell)}_1<\xi_1^{(\ell)}<0$.\\
For $\ell=2$, the above two roots bifurcate into two complex roots, with
$\Re{\Bxi^{(\ell)}}<\Re{\xi^{(\ell)}}$,
$|\Im{\Bxi^{(\ell)}}|>|\Im{\xi^{(\ell)}}|$.\\
Generally, for even $\ell$, there are $\ell$ complex zeros with
\begin{eqnarray}
\Re{\Bxi}^{(\ell)}_k<\Re{\xi}^{(\ell)}_{k},
|\Im{\Bxi}^{(\ell)}_k|>|\Im{\xi}^{(\ell)}_k|,~~k=1,2,\ldots,\ell/2.
\end{eqnarray}
All $\Beta_k^{(\ell,n)}$ are sandwiched between $\Bxi^{(\ell)}_k$
and $\xi^{(\ell)}_k$.  As an even $\ell$ changes to $\ell+1$ which
is odd, all zeros move to the right with the real and the absolute
value of the imaginary parts increased, and a new real zero
appears to the left of all the complex zeros on the negative real
axis. As $\ell$ increases further, the complex zeros move as
described before, and the zero on the negative real axis
bifurcates into two complex zeros, giving an even number of
complex zeros.  These patterns continue as $\ell$ increases.

\begin{figure}[htbp]
\caption{\label{fig6} L2: Distributions of the zeros
$\Beta_k^{(\ell,n)},~\eta_j^{(\ell,n)}$ ($\blacklozenge$),
$\Bxi^{(\ell)}_k$ ($\bigcirc$) and $\xi^{(\ell)}_k $
($\blacksquare$) for the L2 Laguerre polynomials, with $g=2$ and
$n=2$.  The four diagrams correspond to $\ell=1 (a), 2 (b), 3 (c)$
and $20 (d)$, respectively.}
\begin{center}
(a) ~~~ \includegraphics{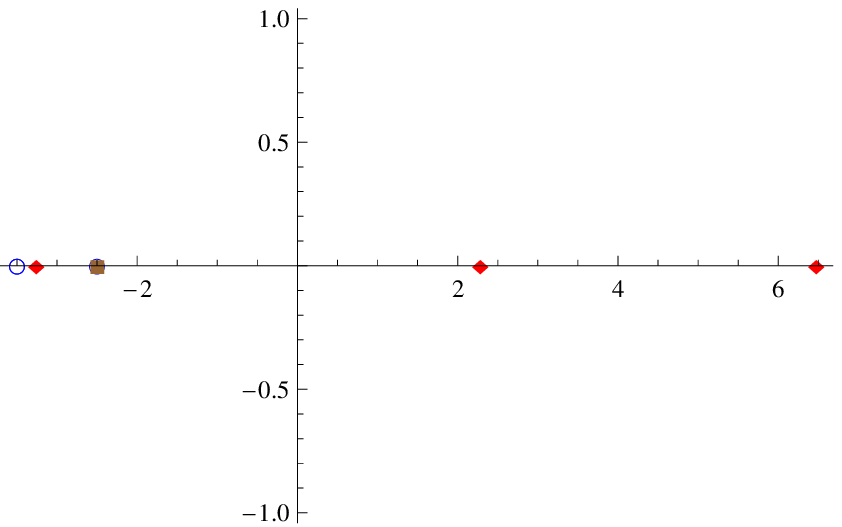}\\~\\
(b) ~~~ \includegraphics{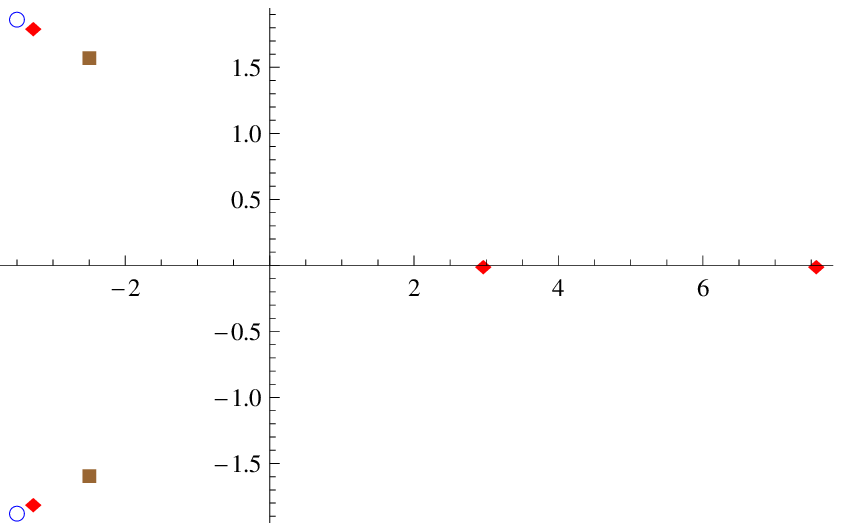}\\~\\
(c) ~~~ \includegraphics{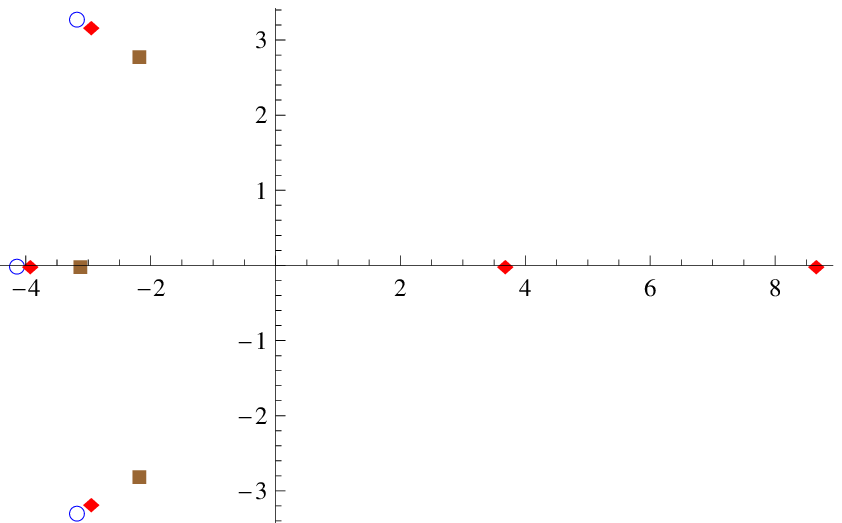}\\~\\
(d) ~~~ \includegraphics{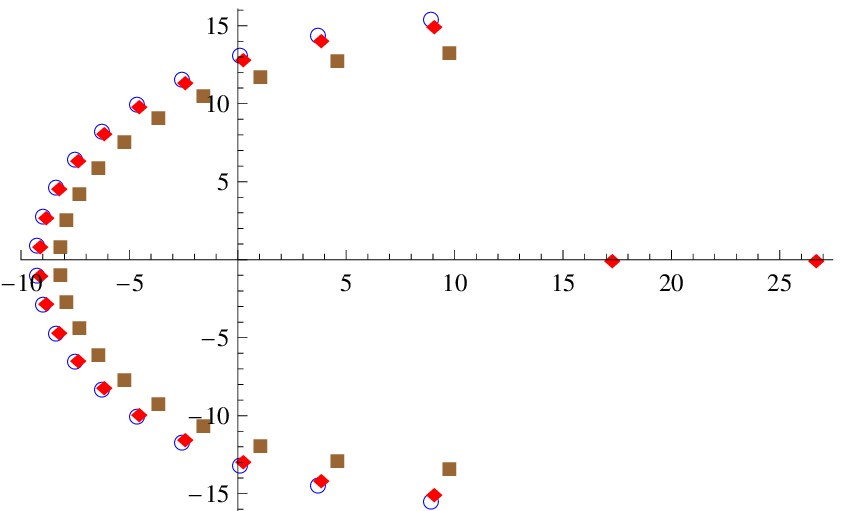}
\end{center}
\end{figure}

\begin{figure}[htbp]
\caption{\label{fig7} L2: Distributions of the zeros
$\Beta_k^{(\ell,n)},~\eta_j^{(\ell,n)}$ ($\blacklozenge$),
$\Bxi^{(\ell)}_k$ ($\bigcirc$) and $\xi^{(\ell)}_k $
($\blacksquare$) for the L2 Laguerre polynomials, with $g=5$ and
$n=2$.  The four diagrams correspond to $\ell=1 (a), 2 (b), 3 (c)$
and $20 (d)$, respectively.}
\begin{center}
(a) ~~~ \includegraphics{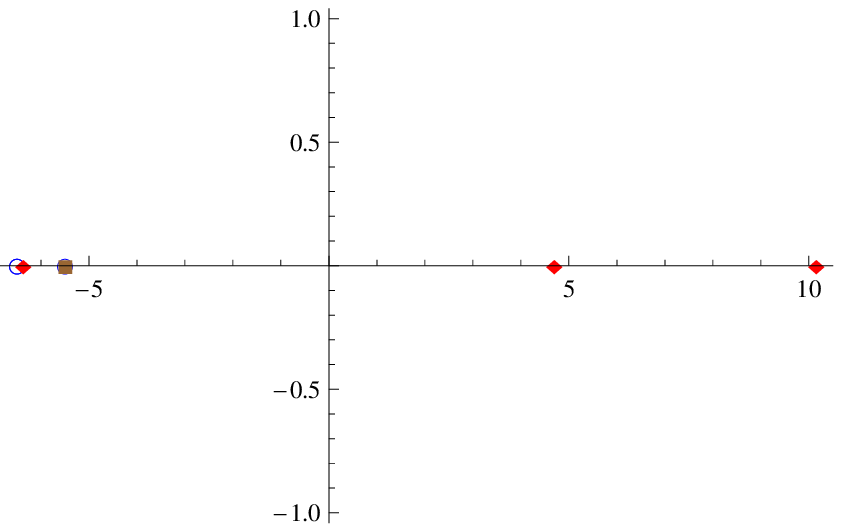}\\~\\
(b) ~~~ \includegraphics{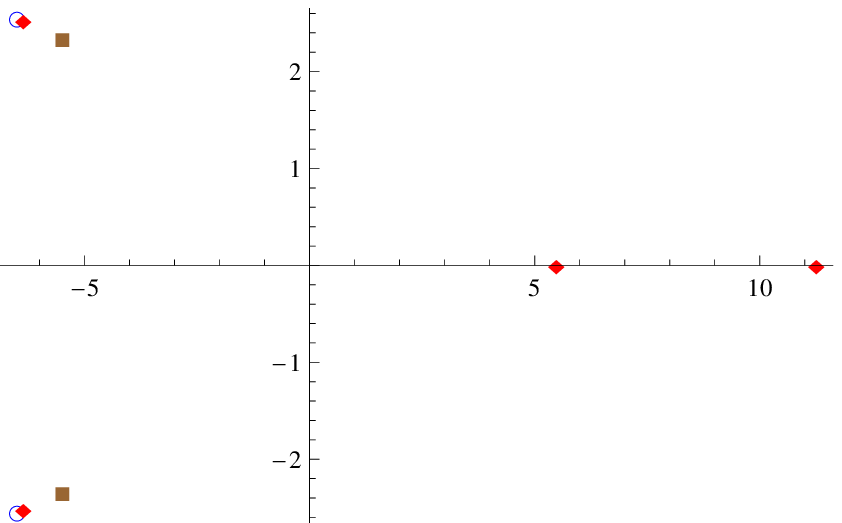}\\~\\
(c) ~~~ \includegraphics{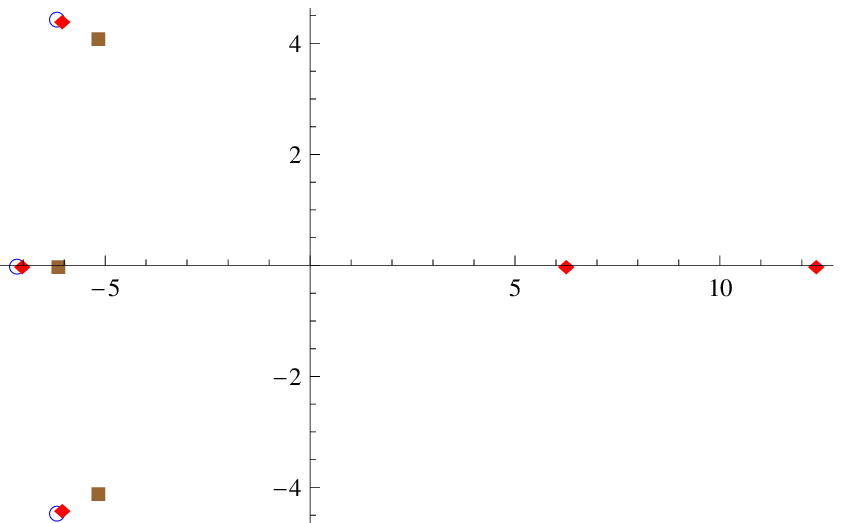}\\~\\
(d) ~~~ \includegraphics{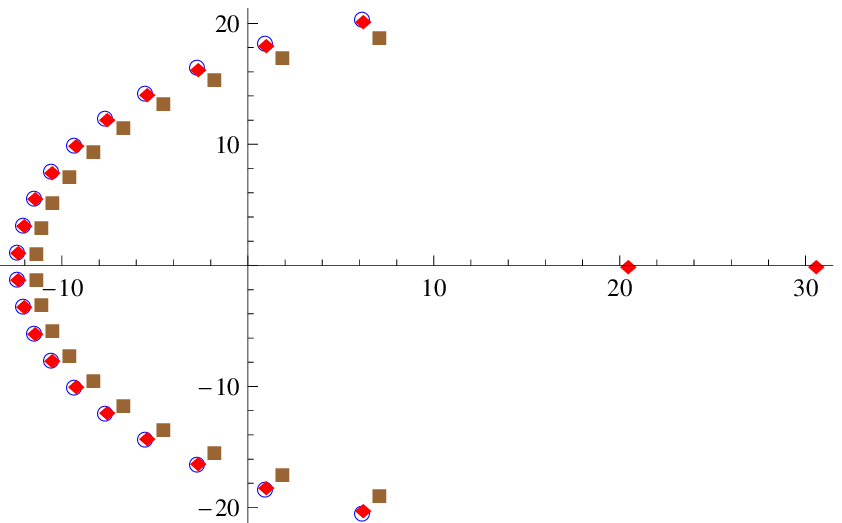}
\end{center}
\end{figure}

Figs.~\ref{fig6} and \ref{fig7} show these behaviors for some
selected parameters.  For large $\ell$, these zeros distribute in
a horse-shoe pattern.

\subsubsection{J2 Jacobi}

For $\ell=1$, there is one real root each for $\xi_\ell(\eta;
g+1)$ and $\xi_\ell(\eta; g)$,
with $\Bxi^{(\ell)}_1>\xi_1^{(\ell)}>1$.\\
For $\ell=2$, the above two roots bifurcate into two complex roots, with
$\Re{\Bxi^{(\ell)}}>\Re{\xi^{(\ell)}}$,
$|\Im{\Bxi^{(\ell)}}|>|\Im{\xi^{(\ell)}}|$.\\
Generally, for even $\ell$, there are $\ell$ complex zeros with
\begin{eqnarray}
\Re{\Bxi}^{(\ell)}_k>\Re{\xi}^{(\ell)}_{k},
|\Im{\Bxi}^{(\ell)}_k|>|\Im{\xi}^{(\ell)}_k|,~~k=1,2,\ldots,\ell/2.
\end{eqnarray}
As $\ell$ changes to $\ell+1$ which is odd, all zeros move toward
the $y$-axis,
with the real parts decreased, and a
new real zero appears to the right of all the complex zeros on the
real $x$-axis.  As $\ell$ increases further, the complex zeros
move as described before, and the zero on the real axis bifurcates
into two complex zeros, giving an even number of complex zeros.  The
absolute value of the imaginary part of the complex zeros may
increase initially, but eventually decrease as $\ell$ increases.
This pattern continues as $\ell$ increases.

\begin{figure}[htbp]
\caption{\label{fig8} J2: Distributions of the zeros
$\Beta_k^{(\ell,n)},~\eta_j^{(\ell,n)}$ ($\blacklozenge$),
$\Bxi^{(\ell)}_k$ ($\bigcirc$) and $\xi^{(\ell)}_k $
($\blacksquare$) for the J2 Laguerre polynomials, with $g=3,  h=4$
and $n=4$.  The four diagrams correspond to $\ell=1 (a), 2 (b), 3
(c)$ and $20 (d)$, respectively.}
\begin{center}
(a) ~~~ \includegraphics{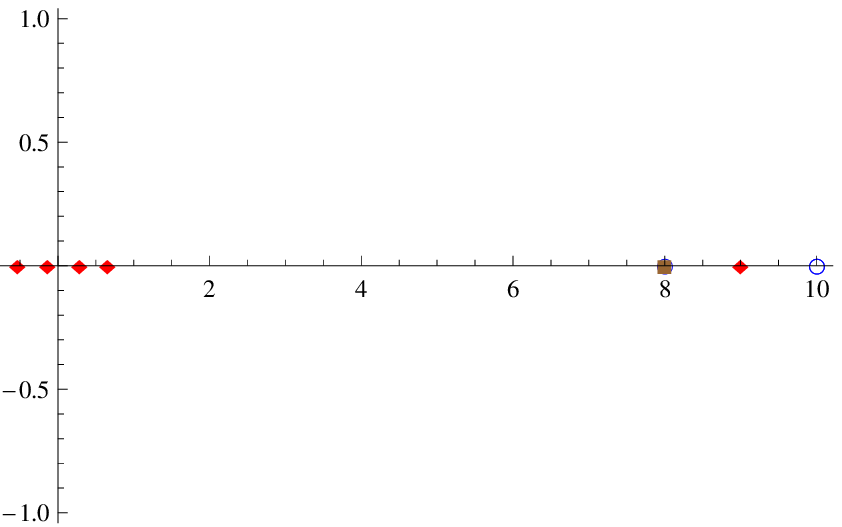}\\~\\
(b) ~~~ \includegraphics{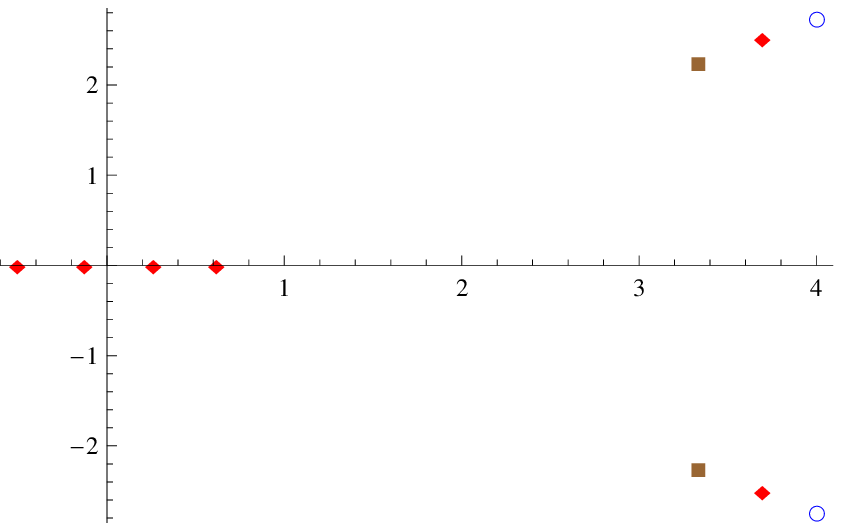}\\~\\
(c) ~~~ \includegraphics{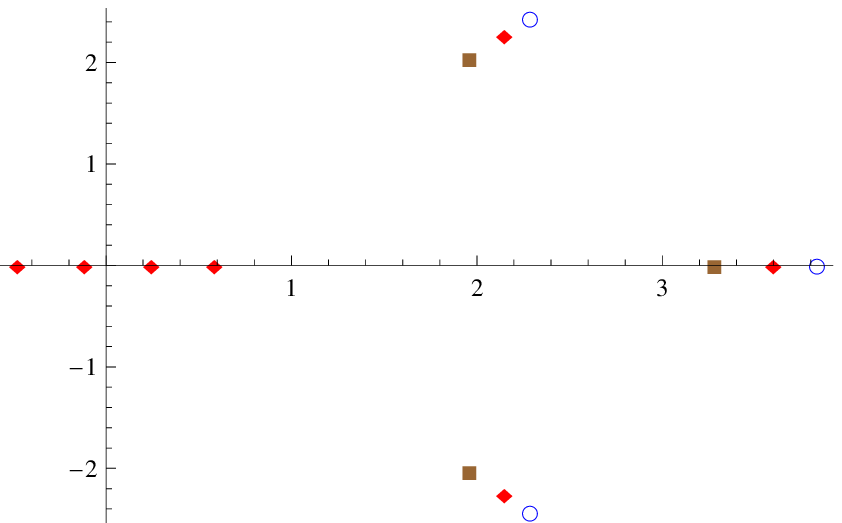}\\~\\
(d) ~~~ \includegraphics{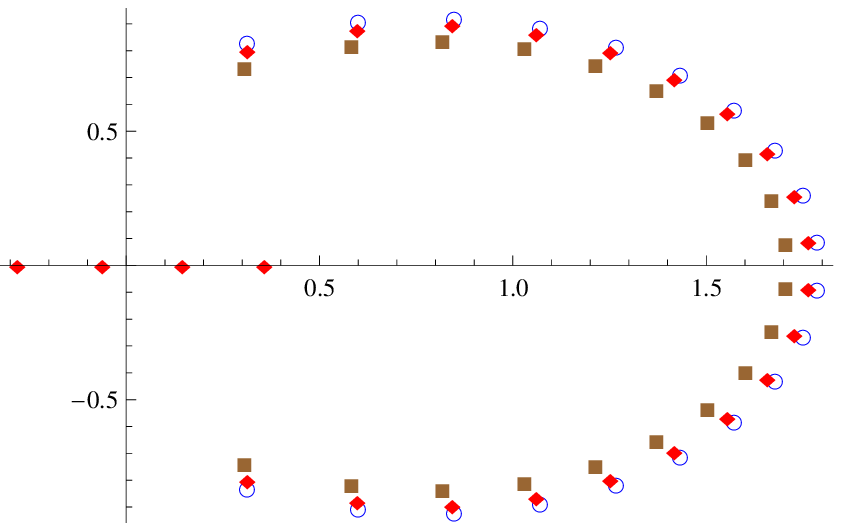}
\end{center}
\end{figure}

\begin{figure}[htbp]
\caption{\label{fig9} J2: Distributions of the zeros
$\Beta_k^{(\ell,n)},~\eta_j^{(\ell,n)}$ ($\blacklozenge$),
$\Bxi^{(\ell)}_k$ ($\bigcirc$) and $\xi^{(\ell)}_k $
($\blacksquare$) for the J2 Laguerre polynomials, with $g=7,  h=8$
and $n=4$.  The four diagrams correspond to $\ell=1 (a), 2 (b), 3
(c)$ and $20 (d)$, respectively.}
\begin{center}
(a) ~~~ \includegraphics{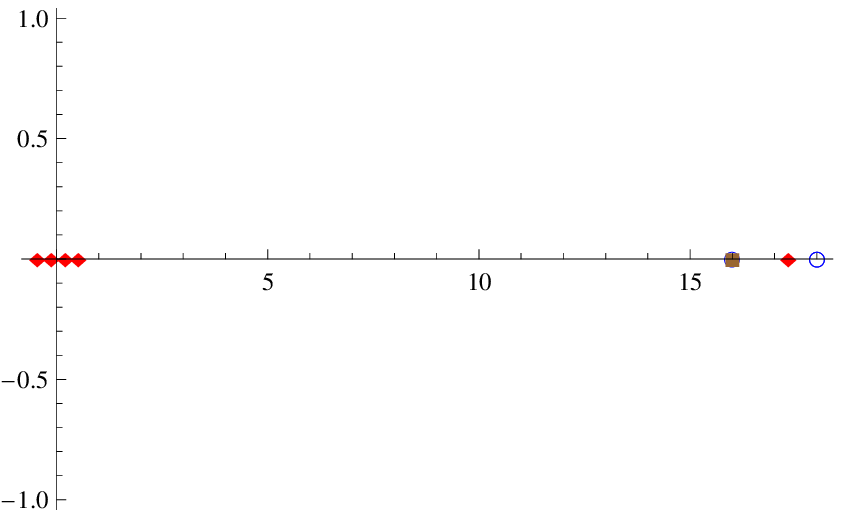}\\~\\
(b) ~~~ \includegraphics{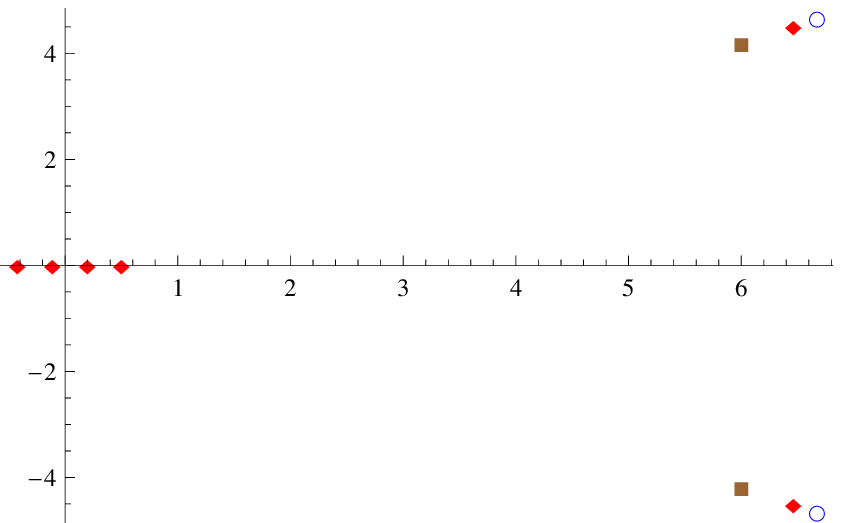}\\~\\
(c) ~~~ \includegraphics{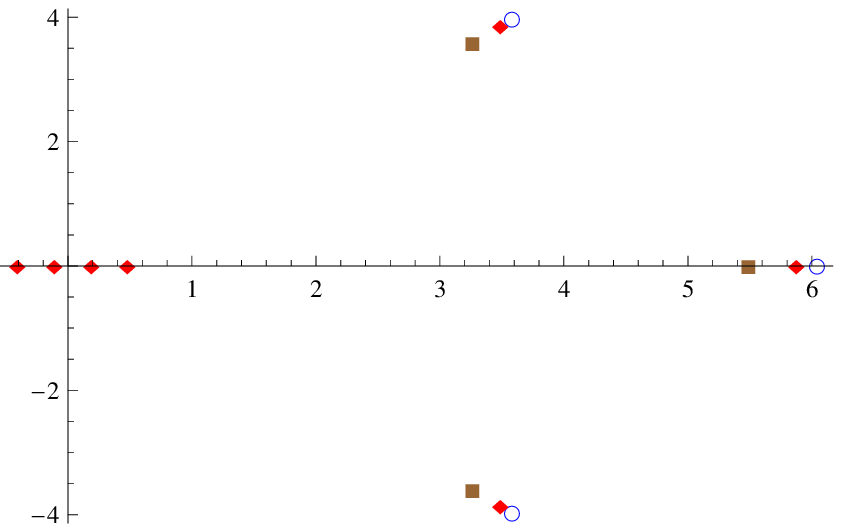}\\~\\
(d) ~~~ \includegraphics{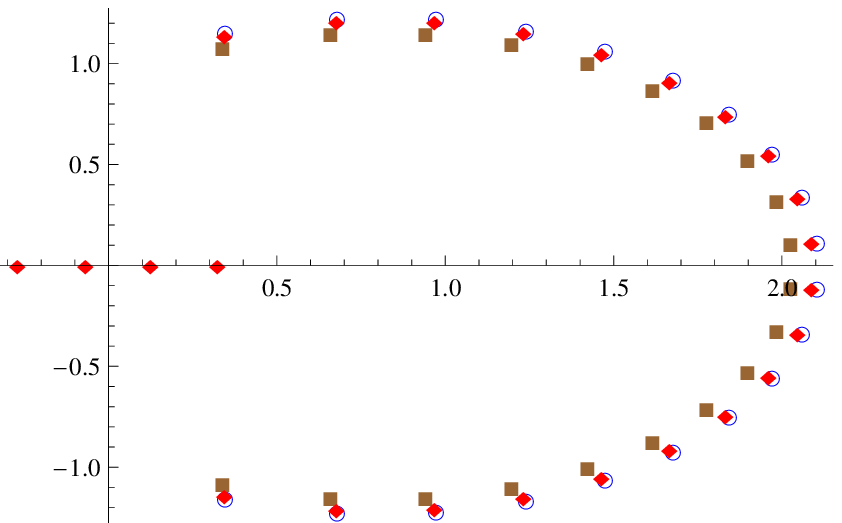}
\end{center}
\end{figure}

Figs.~\ref{fig8} and \ref{fig9} show these behaviors for some
selected parameters. For large $\ell$, these zeros distribute in a
horse-shoe pattern.

\section{Ordinary zeros of $X_{\ell}$ polynomials}
\label{ordinary-zeros}

In the case of $X_{\ell}$ polynomials
$P_{\ell,n}(\eta;\bm{\lambda})$, it has $n$ zeros in the (ordinary) domain where
the weight function is defined, that is $(0,\infty)$ for the L1
and L2 polynomials and $(-1,1)$ for the J1 and J2 polynomials. The
behavior of these zeros are the same as those of other ordinary
orthogonal polynomials.

\subsection{Behaviors as $n$ increases at fixed $\ell$}

This is guaranteed by the oscillation theorem of the
one-dimensional quantum mechanics, since
$P_{\ell,n}(\eta;\bm{\lambda})$ are obtained as the polynomial
part of the eigenfunctions of a shape invariant quantum mechanical
problem.  Explicitly, as $n$ changes to $n+1$, all zeros
$P_{\ell,n}$ decrease, and a new zero appears from the right. Thus
the $n$ zeros of $P_{\ell,n}(\eta;\bm{\lambda})$ and the $n+1$
zeros of $P_{\ell,n+1}(\eta;\bm{\lambda})$ interlace with each
other: each zero of $P_{\ell,n}(\eta;\bm{\lambda})$ is surrounded
by two zeros of $P_{\ell,n+1}(\eta;\bm{\lambda})$.

Figs.~\ref{fig1}-\ref{fig3} show these behaviors for selected
parameters.

\subsection{Behaviors as $\ell$ increases at fixed $n$}

From Figs.~\ref{fig4}-- \ref{fig7}, one sees that for L1 and L2
Laguerre polynomials (whose zeros are positive in the ordinary
domains), all the $n$ zeros shift to the right as $\ell$
increases.

For J2 Jacobi polynomials, the positive (negative) zeros shift
left (right) as $\ell$ increases, i.e., they move toward the
origin $\eta=0$.  This is illustrated in Figs.~\ref{fig8} and
\ref{fig9}.

\subsection{Additional observation for the L1 case}


Using the well-known derivative relation
\begin{equation}
\partial_\eta L_n^{(\alpha)}(\eta)=-L_{n-1}^{(\alpha+1)}(\eta)
\label{partial-L}
\end{equation}
 and
\begin{equation}
L_n^{(\alpha)}(\eta)-L_n^{(\alpha-1)}(\eta)=L_{n-1}^{(\alpha)}(\eta),
\end{equation}
we get
\begin{eqnarray}
P_{\ell,\ell}(\eta;g)&=&
L_\ell^{(g+\ell-\frac12)}(-\eta)L_\ell^{(g+\ell-\frac{3}{2})}(\eta)
+L_\ell^{(g+\ell-\frac12)}(\eta)L_\ell^{(g+\ell-\frac{3}{2})}(-\eta)\\
&&~~~~-L_\ell^{(g+\ell-\frac{3}{2})}(\eta)L_\ell^{(g+\ell-\frac{3}{2})}(-\eta).
\nonumber
\end{eqnarray}
Hence when $n=\ell$, the L1 Laguerre is an even function of
$\eta$, and its zeros are symmetric w.r.t $\eta=0$.

\section{Proof that $\Beta^{(\ell,n)}_k\to \xi^{(\ell)}_k$ as $n\to \infty$}
\label{sect:eta-flow}

As mentioned before, for $n=0$, we have $\Beta^{(\ell,0)}_k=
\Bxi^{(\ell)}_k$, as $P_{n=0}=1$.
We shall show that as $n\to \infty$, $\Beta^{(\ell,n)}_k\to
\xi^{(\ell)}_k$.  This amounts to showing that in this limit,
$\partial_{\eta} P_n$ dominates over $P_n$

\subsection{L1 and L2 cases}

We shall make use of   the above derivative relation
\eqref{partial-L} and (Perron) Theorem 8.22.3 of \cite{Szego},
namely,
\begin{equation}
L_n^{(\alpha)}(\eta)\cong\frac{e^{\frac{\eta}{2}}}{2\sqrt{\pi}}(-\eta)^{-\frac{\alpha}{2}-\frac{1}{4}}
n^{\frac{\alpha}{2}-\frac{1}{4}} e^{2\sqrt{-n\eta}},~~\alpha\in
\mathbb{R},~~\eta\in\mathbb{C}\backslash (0,\infty),
\end{equation}
which gives the asymptotic form of $L_n^{(\alpha)}(\eta)$ for large
$n$.
For the L1 and L2 cases, we have $\alpha=g+\ell-3/2$ and
$g+\ell+1/2$, respectively.

One finds
\begin{equation}
\left|\frac{L_n^{(\alpha)}(\eta)}{\partial_\eta
L_n^{(\alpha)}(\eta)}\right|\sim
\left|-\frac{1}{\sqrt{n}}(-\eta)^\frac12\right|.
\end{equation}

For large $n$ with fixed $\eta$, $\partial_\eta
L_n^{(\alpha)}(\eta)$ dominates over $L_n^{(\alpha)}(\eta)$, and
thus the zeros of $P_{\ell,n}$ are determined by those of
$\xi_\ell(\eta;g)$ as $n\to \infty$.

\subsection{J2 Jacobi}

For the asymptotic form of $P_n^{(\alpha,\beta)}(\eta)$ for large
$n$, we shall make use of Theorem 8.21.7 of \cite{Szego}:
\begin{eqnarray}
P_n^{(\alpha,\beta)}(\eta) &\cong&
(\eta-1)^{-\frac{\alpha}{2}}(\eta+1)^{-\frac{\beta}{2}}
\{\sqrt{\eta+1}+\sqrt{\eta-1}\}^{\alpha+\beta} \nonumber\\
&&\times \frac{(\eta^2-1)^{-1/4}}{\sqrt{2\pi n}}
\{\eta+\sqrt{\eta^2-1}\}^{n+\frac12},\\
&&~~~~~~~~~~~~~~~~~~\alpha,~\beta\in \mathbb{R},
~~\eta\in\mathbb{C}\backslash \left[-1,1\right], \nonumber
\end{eqnarray}
and
\begin{equation}
\partial_\eta P_n^{(\alpha,\beta)}(\eta)
 =\tfrac12(n+\alpha+\beta+1)P_{n-1}^{(\alpha+1,\beta+1)}(\eta).
\end{equation}

One finds
\begin{equation}
\frac{P_n^{(\alpha,\beta)}(\eta)}{\partial_\eta
P_n^{(\alpha,\beta)}(\eta)}\sim \frac{2}{(n+\alpha+\beta
+1)}\sqrt{\frac{n-1}{n}}\left(\eta^2-1\right)^\frac12\frac{\eta+\sqrt{\eta^2-1}}
{(\sqrt{\eta+1}+\sqrt{\eta-1})^2}.
\end{equation}

Again, for large $n$ with fixed $\eta$, $\partial_\eta
P_n^{(\alpha,\beta)}(\eta)$ dominates over
$P_n^{(\alpha,\beta)}(\eta)$, and thus the zeros of $P_{\ell,n}$
are determined by those of $\xi_\ell(\eta;g)$ as $n\to \infty$.

\section{Behaviors at large $g$ and /or $h$}

\subsection{L1 Laguerre}

As $g$ increases, we have
\begin{eqnarray}
|\Bxi^{(\ell)}_k|,~~|\xi^{(\ell)}_k|,~~|\Beta_k^{(\ell,n)}|,~~
|\eta_k^{(\ell,n)}|
\end{eqnarray}
all increase. That is, all the zeros move away from the $y$-axis.
This can be seen from Figs.~\ref{fig4} and \ref{fig5}.

In fact for large $g$, we have $\xi_\ell(\eta;g+1)\approx
\xi_\ell(\eta;g)$. Hence
\begin{eqnarray}
P_{\ell,n}(\eta;g)&\approx & \xi_\ell(\eta;g)
\left[L_n^{(g+\ell-\frac{3}{2})}(\eta)-\partial_\eta
L_n^{(g+\ell-\frac{3}{2})}(\eta)\right]\nonumber\\
&\approx & \xi_\ell(\eta;g) L_n^{(g+\ell-\frac12)}(\eta).
\end{eqnarray}
For $g\gg 1$, $P_{\ell,n}(\eta;g)$ approaches
\begin{eqnarray}
P_{\ell,n}(\eta;g)\approx L_\ell^{(g+\ell)}(-\eta)
L_n^{(g+\ell)}(\eta).
\label{factL1}
\end{eqnarray}
Thus the extra ($\Beta_k^{(\ell,n)}$) and the ordinary
($\eta_k^{(\ell,n)}$) zeros of $P_{\ell,n}(\eta;g)$ are given by
the zeros of $L_\ell^{(g+\ell)}(-\eta)$ and
$L_n^{(g+\ell)}(\eta)$, respectively.

\subsection{L2 Laguerre}

As $g$ increases, we have $\Re\Bxi^{(\ell)}_k,~~\Re\xi^{(\ell)}_k$
decreased, $|\Im\Bxi^{(\ell)}_k|,~~|\Im\xi^{(\ell)}_k|$ increased,
and $\eta^{(\ell,n)}_k$ increased. This is easily seen from
Figs.~\ref{fig6} and \ref{fig7}. That is, the zeros
$\Bxi^{(\ell)}_k,~\xi^{(\ell)}_k$, and hence $\Beta^{(\ell,n)}_k$,
all are moving leftwards and  away from the $x$-axis, while the
ordinary zeros $\eta^{(\ell,n)}_k$ are moving towards the right.

In fact for large $g$, we have $\xi_\ell(\eta;g+1)\approx
\xi_\ell(\eta;g)$. Hence
\begin{eqnarray}
P_{\ell,n}(\eta;g)&\approx & \xi_\ell(\eta;g)
\left[\left(g+\frac12\right)L_n^{\left(g+\ell+\frac12\right)}(\eta)+\eta\partial_\eta
L_n^{\left(g+\ell+\frac12\right)}(\eta)\right].
\end{eqnarray}
Using Eqs.~(E.2), (E.10) and (E.9) of \cite{HOS}, we arrive at
\begin{eqnarray}
P_{\ell,n}(\eta;g)&\approx & \xi_\ell(\eta;g)
\left[\left(g+\ell+\frac12+n\right)L_n^{\left(g+\ell-\frac12\right)}(\eta)-\ell
L_n^{\left(g+\ell+\frac12\right)}(\eta)\right].
\end{eqnarray}
For $g\gg 1$, $P_{\ell,n}(\eta;g)$ approaches
\begin{eqnarray}
P_{\ell,n}(\eta;g)\approx L_\ell^{(-g-\ell)}(\eta)
L_n^{(g+\ell)}(\eta).
\label{factL2}
\end{eqnarray}
Thus the extra ($\Beta_k^{(\ell,n)}$) and the ordinary
($\eta_k^{(\ell,n)}$) zeros of $P_{\ell,n}(\eta;g)$ are given by
the zeros of $L_\ell^{(-g-\ell)}(\eta)$ and
$L_n^{(g+\ell)}(\eta)$, respectively.

\subsection{J2 Jacobi}

As $g,~h$ increases, we have
$\Re\Bxi^{(\ell)}_k,~~\Re\xi^{(\ell)}_k,~
|\Im\Bxi^{(\ell)}_k|,~~|\Im\xi^{(\ell)}_k|$ increased, as is
evident from Figs.~\ref{fig8} and \ref{fig9}. The extra zeros
$\Beta^{(\ell,n)}_k$, being in between these zeros, follow the
same pattern. That is, the zeros
$\Bxi^{(\ell)}_k,~\xi^{(\ell)}_k$, and hence $\Beta^{(\ell,n)}_k$,
all are moving away from the $x$ and $y$-axes. The ordinary zeros
$\eta^{(\ell,n)}_k$ will have their norm
$|\eta^{(\ell,n)}_k|$decrease in general as $g$ increases. Thus
these zeros move towards the $y$-axis.

In fact for large $g$ and $h$, we have ($\alpha\equiv
g+\ell+\frac12,~\beta\equiv h+\ell-\frac{3}{2}$)
\begin{eqnarray}
P_{\ell,n}(\eta;g,h)&\approx & \xi_\ell(\eta;g,h)
\left[\left(g+\frac12\right)P_n^{(\alpha,\beta)}(\eta)-\left(1-\eta\right)\partial_\eta
P_n^{(\alpha,\beta)}(\eta)\right].
\end{eqnarray}
Using Eqs.~(E.13) and (E.23) of \cite{HOS}, we arrive at
\begin{eqnarray}
P_{\ell,n}(\eta;g,h)&\approx & \xi_\ell(\eta;g,h)
\left[\left(\alpha+n\right)P_n^{(\alpha +1,\beta+1)}(\eta)-\ell
P_n^{(\alpha,\beta)}(\eta)\right].
\end{eqnarray}
For $g\gg 1$ and $h\gg 1$, $P_{\ell,n}(\eta;g,h)$ approaches
\begin{eqnarray}
P_{\ell,n}(\eta;g)\approx P_\ell^{(-g-\ell,h+\ell)}(\eta)
P_n^{(g+\ell,h+\ell)}(\eta).
\label{factJ2}
\end{eqnarray}
Thus the extra ($\Beta_k^{(\ell,n)}$) and the ordinary
($\eta_k^{(\ell,n)}$) zeros of $P_{\ell,n}(\eta;g,h)$ are given by
the zeros of $P_\ell^{(-g-\ell,h+\ell)}(\eta)$ and $P_n^{(g+\ell,
h+\ell)}(\eta)$, respectively.

\subsubsection{Additional observation: $h\gg g$}

\begin{figure}[htbp]
\caption{\label{fig10} J2: Distributions of the zeros
$\Beta_k^{(\ell,n)},~\eta_j^{(\ell,n)}$ ($\blacklozenge$),
$\Bxi^{(\ell)}_k$ ($\bigcirc$) and $\xi^{(\ell)}_k $
($\blacksquare$) for the J2 Laguerre polynomials, with $g=2$ ,
$\ell=10$ and $n=4$. The two diagrams correspond to $h=50 (a)$ and
$100 (b)$, respectively.}
\begin{center}
(a) ~~~ \includegraphics{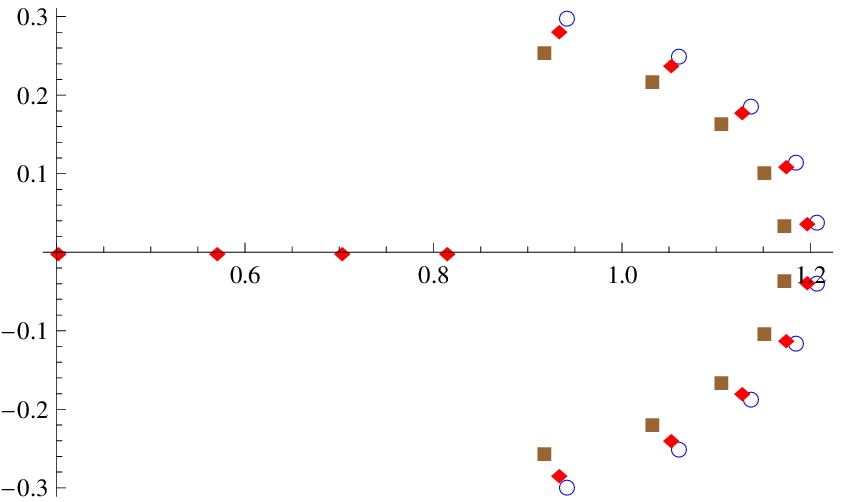}\\~\\
(b) ~~~ \includegraphics{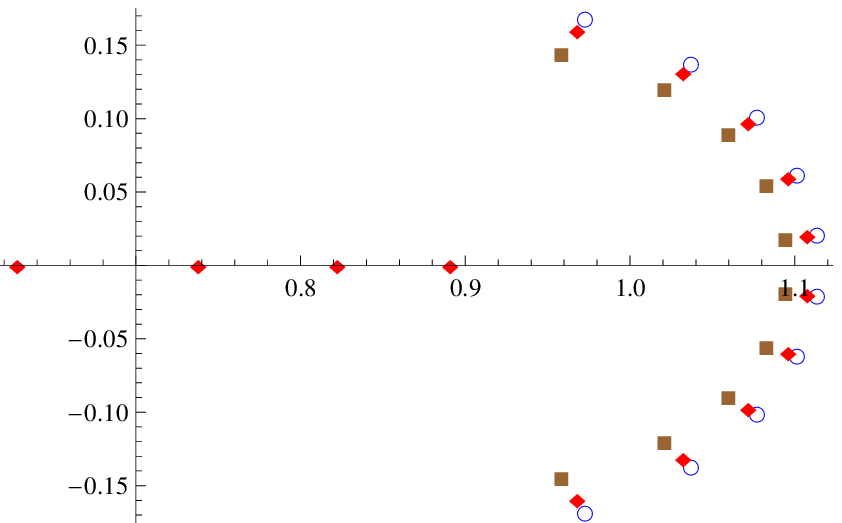}
\end{center}
\end{figure}

For $h\gg g$, all zeros, i.e., $\Bxi_j^{(\ell)},~\xi_j^{(\ell)},~\Beta_k^{(\ell,n)},~\eta_k^{(\ell,n)}$,
gather around $\eta=1$. This can be understood as follows. From
the series expansion of the Jacobi polynomials, Eq.~(E.11)  of
\cite{HOS},
\begin{equation}
 P_n^{(\alpha,\beta)}(\eta)=\frac{(\alpha+1)_n}{n!}
 \sum_{k=0}^n\frac{1}{k!}\frac{(-n)_k(n+\alpha+\beta+1)_k}{(\alpha+1)_k}
 \Bigl(\frac{1-\eta}{2}\Bigr)^k,
 \label{Jexp}
\end{equation}
one sees that, for $h\gg g$, the absolute value of
$P_{\ell,n}(\eta;g,h)$ is large near $\eta=-1$ and small at
$\eta=1$. Hence, in this limit, the zeros of
$P_{\ell,n}(\eta;g,h)$ distribute very near $\eta=1$. We show this
in Fig.~\ref{fig10} for certain parameters.

\section{Summary}

The discovery of new types of orthogonal polynomials, called the
exceptional $X_\ell$ Laguerre and Jacobi polynomials has aroused
great interest in the last two years. Unlike the well-known
classical orthogonal polynomials which start with constant terms,
these new polynomials $P_{\ell,n}(\eta;\bm{\lambda})$ have the lowest
degree $\ell=1,2,\ldots$, and yet they form a complete set with
respect to some positive-definite measure. Many essential
properties have been studied in \cite{HOS}.

In this paper, we have considered the distributions of the zeros
of these new polynomials as some parameters of the Hamiltonians
change. The $X_{\ell}$ polynomials $P_{\ell,n}(\eta;\bm{\lambda})$
has $n$ zeros in the ordinary domain where the weight function is
defined, that is $(0,\infty)$ for the L1 and L2 polynomials and
$(-1,1)$ for the J1 and J2 polynomials. The behavior of these
ordinary zeros are the same as those of other ordinary orthogonal
polynomials.  In addition to these $n$ zeros, there are extra
$\ell$ zeros outside the ordinary domain.

For the ordinary zeros, their distribution as $n$ increases at a
fixed $\ell$ follows the patterns of the zeros of the ordinary
classical orthogonal polynomials: they are governed by the
oscillation theorem, and the $n+1$ zeros of
$P_{\ell,n+1}(\eta;\bm{\lambda})$ interlace with the $n$ zeros of
$P_{\ell,n}(\eta;\bm{\lambda})$.  On the other hand, when $\ell$
increases at a fixed $n$, the type L1 and L2 Laguerre polynomials
will have all their $n$ zeros shifted to the right. For the J1 and
the J2 Jacobi polynomials, both the positive and negative zeros
move toward the origin $\eta=0$ as $\ell$ increases.

For the $\ell$ extra zeros of $P_{\ell,n}(\eta;\bm{\lambda})$,
each and everyone of them is sandwiched between the corresponding
zeros of the deforming polynomials
$\xi_{\ell}(\eta;\bm{\lambda}+\bm{\delta})$ and
$\xi_{\ell}(\eta;\bm{\lambda})$.  As $n$ increases at a fixed
$\ell$, the extra zeros move from the zeros of
$\xi_{\ell}(\eta;\bm{\lambda}+\bm{\delta})$ to those of
$\xi_{\ell}(\eta;\bm{\lambda})$.

The behaviors of the extra zeros as $\ell$ increases at a fixed
$n$ are more complex.  For the L1 the Laguerre polynomials, all
its extra zeros lie on the negative $x$-axis. So as $\ell$
increases by one, the number of the extra zeros increases from
$\ell$ to $\ell+1$. For the L2 Laguerre and J1 and J2 Jacobi
polynomials, they have $\ell/2$ pairs of complex zeros for even
$\ell$, and $(\ell-1)/2$ pairs of complex zeros and a real zero
outside the ordinary domains where the weight functions are
defined. As $\ell$ increases, all the complex zeros move toward
the right in the case of the L2 Laguerre and J1 Jacobi
polynomials, and toward the left for the J2 Jacobi polynomials,
while the extra real zeros bifurcate into new pairs of complex
zeros. For large $\ell$, these zeros appear to distribute
symmetrically with respect to the $x$-axis in horse-shoe patterns.
It is interesting to note that in the asymptotic regions of the
parameters ($g\gg1$,  $h\gg 1$), the exceptional polynomial
$P_{\ell,n}(\eta,\bm{\lambda})$ is expressed as the product of the
original polynomial $P_n(\eta)$ and the deforming polynomial
$\xi_\ell(\eta;\bm{\lambda})$, \eqref{factL1}, \eqref{factL2} and
\eqref{factJ2}.

\section*{Acknowledgments}

This work is supported in part by the National Science Council
(NSC) of the Republic of China under Grant NSC
NSC-99-2112-M-032-002-MY3 (CLH), and in part by Grants-in-Aid for
Scientific Research from the Ministry of Education, Culture,
Sports, Science and Technology, No.19540179 (RS). RS wishes to
thank the R.O.C.'s National Center for Theoretical Sciences and
National Taiwan University for the hospitality extended to him
during his visit in which part of the work was done.

\newpage

\newpage

\newpage

\newpage

\newpage

\newpage

\end{document}